# Identifying Economic Factors Affecting Unemployment Rates in the United States

Alrick Green[1], Ayesha Nasim[1], Jaydeep Radadia[1], Devi Manaswi Kallam[1], Viswas Kalyanam[1], Samfred Owenga[1,] and Huthaifa I. Ashqar[2,*]

[1] Department of Professional Studies, University of Maryland, Baltimore County, MD, USA
[2] Arab American University, Jenin, Palestine

[*] Corresponding author (hiashqar@vt.edu)

## Abstract

In this study, we seek to understand how macroeconomic factors such as GDP, inflation, Unemployment Insurance, and S&P 500 index; as well as microeconomic factors such as health, race, and educational attainment impacted the unemployment rate for about 20 years in the United States. Our research question is to identify which factor(s) contributed the most to the unemployment rate surge using linear regression. Results from our studies showed that GDP (negative), inflation (positive), Unemployment Insurance (contrary to popular opinion; negative), and S&P 500 index (negative) were all significant factors, with inflation being the most important one. As for health issue factors, our model produced resultant correlation scores for occurrences of Cardiovascular Disease, Neurological Disease, and Interpersonal Violence with unemployment. Race as a factor showed a huge discrepancies in the unemployment rate between Black Americans compared to their counterparts. Asians had the lowest unemployment rate throughout the years. As for education attainment, results showed that having a higher education attainment significantly reduced one's chance of unemployment. People with higher degrees had the lowest unemployment rate. Results of this study will be beneficial for policymakers and researchers in understanding the unemployment rate during the pandemic.

## Introduction

Unemployment is a serious economic issue that affects almost all countries and all people either directly or indirectly. In most market economies, not everyone will have jobs all the time. It is important to understand unemployment and its causes in order to develop policies to help those individuals who suffer from the adverse effects of being unemployed. The unemployment rate in the United States has never been zero and usually remained between 5 and 6 percent with some years being slightly better or worse. Unemployment rates fluctuate everywhere depending on the economic circumstances of the time. The Bureau of labor statistics considers a person unemployed when someone is laid off from the work for the reasons except misconduct, looking for a job for more than four weeks, and meets the above conditions and is ready to work (BLS.gov (2020)).

On the macroeconomic level, there are a multitude of factors affecting unemployment and the economics fraternity agree on them. Van Ours et al. have worked on the relationship between unemployment duration and unemployment level (Van Ours, J. C. and Vodopivec, M. (2006)). They proposed the disincentivizing effects of unemployment insurance benefits on the unemployment rate. Digvijay, D. B. et al. (2021) found out that the GDP and the unemployment has a negative correlation. As the productivity of the nation increases, unemployment shows a downward trend. Shegay et al identified several factors affecting employment not limited to economic factors (Shegay, O., Tatyana, E., and Anika, C. (2015)). They used linear regression to find the relationship between the rate of unemployment and economic variables. They found economic, demographic, and educational factors. Among the economic variables were the GDP and the market index. Nobel laureate Milton Friedman advanced a natural state hypothesis to correct the then dominating theory known as "Philip Curve" (Friedman, 1977). He emphasized on the role of inflation in the unemployment level. He argued that in the long run there is no trade-off between the inflation and the unemployment which was contrary to the conventional wisdom.

Although unemployment is strongly determined by the performance of the economy, there are several factors that affect it besides the macroeconomic variables. Some companies may have underutilized employees on their payrolls since laying off workers when demand drops and rehiring them when demand rises costs money (Nichols, A., Mitchell, J., & Lindner, S. 2020). As a result, businesses may be able to raise output to meet increased demand without having to hire additional workers at the outset of a recovery by raising the productivity of their existing workforce. As an economic expansion continues, output growth will be determined by the combined rates of growth in the labor supply and labor productivity. As long as growth in real gross domestic product (GDP) outpaces growth in labor productivity, employment will rise. If employment growth outpaces labor force growth, the unemployment rate will fall (Sprague, S. 2014).

**Burden of Disease**
We reviewed the existing work on the factors contributing to unemployment other than the well-known economic factors, these factors were mainly related to health (Dooley, D., Fielding, J., & Levi, L. (1996)). For our study we selected three factors namely Cardiovascular Disease (CVD), Neurological Disorders, and Interpersonal Violence. From previous studies, by the Australian Government, 82% people with no CVD contribute to the labor force where 64% with heart condition do the same. Similarly, for Mental Disorders it is 80% without while only 39% with the condition (Laplagne, P., Glover, M., & Shomos, A. (2007). Furthermore, According to Legal Momentum, an advocacy group in the United States, the victims of domestic violence lose an average of 137 hours of work a year. Intimate partner violence causes victims to lose the equivalent of 32,000 full-time jobs each year (Workplace Fairness, 2021).

A person's health issues may cause him or her to leave the workforce as the marginal utility of unemployment grows. For a manual worker, more labor productivity, higher remuneration, and a higher possibility of remaining in the labor market are all likely to be linked to better health and well-being. In this sense, engagement in the labor force may be harmful to one's health. For people who are sick, lower predicted salaries minimize the cost of not working (Laplagne, P., Glover, M., & Shomos, A. 2007).

**Race**

The idea of community differentiation study is to explain how various groups interact to create imbalance. In order to get weight on these problems, this section explores how race and unemployment spells react to form a basic economic result: the ability to find job (Browne and Misra 2003). An idea is that such qualities will just be recognized as different groups of people that will join in a logical, positive method (Karren and Sherman 2012). According to the findings, Hispanics saw the greatest increase in unemployment during the COVID-19 pandemic because they had the highest coefficient. This supports prior findings of racial disparities in unemployment rates, with Hispanics being the most influenced by both gender and occupation. However, in terms of unemployment, the Asian community continues to outperform the White population. This could be due to the different types of jobs that each race/ethnicity primarily performs. Asians account for 14.1 percent of service occupations, the hardest damaged by the pandemic, while other minorities, such as Black and Hispanic, account for 20-22 percent of the occupation (Lin, J., M., Loh, A., & A. 2021).

**Education**

Educational attainment plays a critical role in the productivity and the growth of the economy in society. Obtaining an education helps to enhance opportunities to achieve suitable jobs and leads to positive contributions such as making one more competitive in highly specialized careers fields, higher income, financial stability, greater economic mobility, and lowers the risk of unemployment. Studies such as (Card, 2001; Grossman, 2005; Oreopoulos and Salvanes, 2009) investigated the impacts of education on the labor market. Results from these studies showed that education positively impacts the labor market's success economically (earnings, and employment) and non-economically (health, participation, and decrease in criminal activity). Riddell and Song (2011) investigated the influence of education on the transition between unemployment and employment. This study was in hopes to understand whether education improves re employment of unemployed workers and how educational attainment improves the capabilities of the labor force adjustment to economic crisis. Results showed that education significantly increases re-employment success for unemployed workers (Riddell and Song 2011).

Nonetheless, the 2019 COVID economic crisis was an anomalous and unexpected event that severely halted the global economy for over 18 months. Due to its nature of being a health crisis, we seek to understand how our selected macroeconomic factors, and non-economic factors such as health issues, race and ethnicities, and educational attainment impacted the unemployment rate during COVID economic decline and recovery periods. Our research question is to identify which factor(s) contributed the most to the unemployment rate surge. The paper is organized as follows. Next, we have the datasets section that describes the data used for this study. The methodology section outlines our methods used to formulate our analysis. This includes our model framework of OLS linear regression. Next, we have our results section followed by a brief conclusion section.

## Datasets

The data was collected from multiple United Sates federal websites from the year 2000 to 2021. Unemployment rate and Unemployment Insurance was extracted from the Bureau of Labor Statistics. The GDP and the inflation data was collected from the Bureau of Economic Analysis. The nominal GDP was used as the proxy for the GDP data. Since inflation is reflected by the change in the prices of basket of goods, the chained Consumer Price Index was used as a proxy. For the unemployment benefits variable, the average duration of people receiving the unemployment benefits in a month was used. The S&P 500 data was pulled out from Yahoo finance.

Since the GDP data is published on a quarterly basis, we had to convert other variables into quarterly figures. For the marker index, S&P 500 was selected. Since, S&P 500 represents a large proportion of corporations in the United States, it was chosen using the law of large numbers. For the S&P 500, the closing value of the index at the end of each quarter was considered. For the inflation, the average value of the inflation over the span of three months was used. Similarly, for the unemployment insurance and the unemployment rate average quarterly data was used.

The data for the disease prevalence was collected as part of the Global Burden of Disease (GBD) project. It is published by both the researchers at the Institute of Health Metrics and Evaluation (IHME) and the 'Disease Burden Unit' at the World Health Organization (WHO), which was created in 1998. The IHME continues the work that was started in the early 1990s and publishes the Global Burden of Disease study. The data retrieved for this study was for the United States for the period 1990-2016, and out of 24 features available, we selected CVD, Neurological disorders, and Interpersonal violence (per 100,000 persons). However, the datasets for educational attainment, and race compared to unemployment rates were taken from the United States Bureau of Labor Statistics. There was no need for data cleaning and interpolation since the datasets were perfectly available without missing data.

## Methodology

Exploratory Data Analysis (EDA) was performed on the datasets where distribution of the data was analyzed as well as for the outliers. We used jointplot function of python's seasborn library to develop visualizations to illustrated Probability Density Functions (PDF) using Kernel Density Estimate (KDE) plots, scatter plots to display the data points and linear regression for line fitting. From the visualizations, we could see that none of the features' distribution were normal in nature. Furthermore, a positive relationship was observed between unemployment insurance and unemployment rate whereas negative between S&P 500 index and the same.

The box plots showed that outliers could be detected for unemployment insurance feature only.

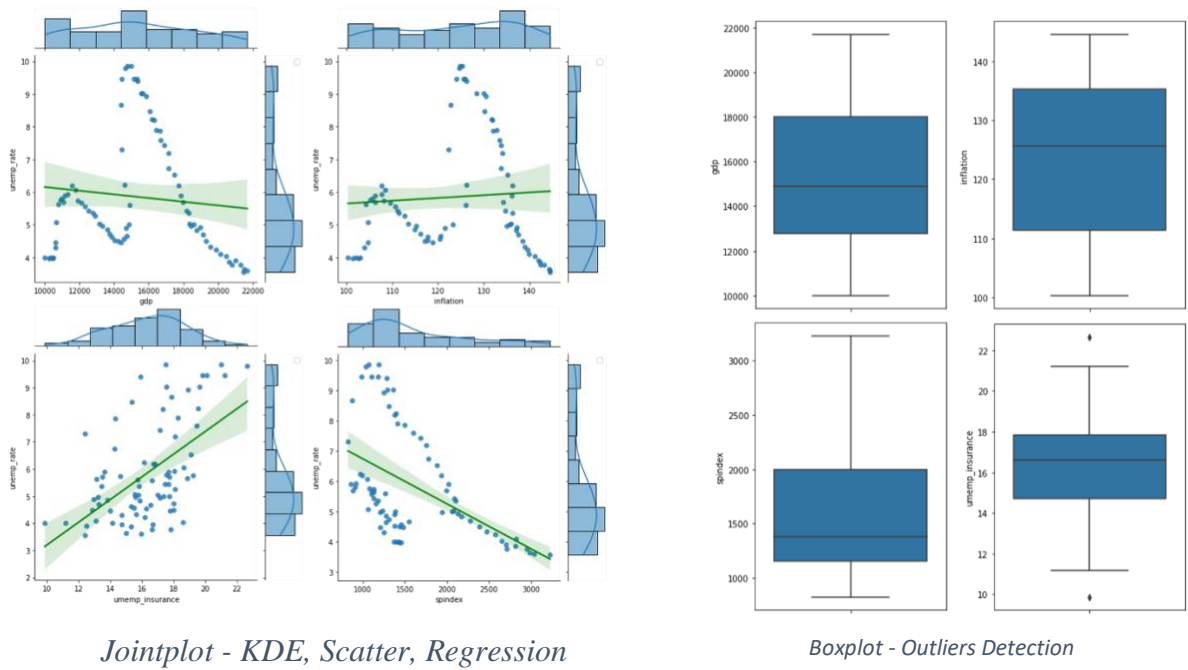

*Jointplot - KDE, Scatter, Regression*     Boxplot - Outliers Detection

Figure 1 Jointplot and Boxplot Illustrating Descriptive Statistical Analysis

We merged the datasets, was split into training and testing datasets to avoid overfitting and allow for validation. After that, pre-processed the data by passing the data through a pipeline which consisted of these steps; outlier removal using Isolation Forest, imputation using Simple Imputer by filling out missing values with the distribution mean, Min Max Scaler, final step of this pipeline had the regression algorithm where we trained the model. We experimented with Linear, Lasso, Ridge, Elastic Net Regression and tree-based Random Forecast algorithms and out of these Random Forest gave the best performance results. The performance of the model is evaluated using certain metrics, we used R-Squared for model selection which is commonly used to explain the amount of variation in the data explained by the model. The higher the value, the better the fit. Using our data model, we ultimately deduced the impact of individual features on the unemployment rate.

For regression analysis on Microecomic factors, due to a smaller dataset we used Linear Regression. It is a simple yet powerful technique to determine the correlation between an independent variable and a dependent variable. However, the multi-linear regression analysis attempts to account for the variation in the dependent variable based on the multiple independent variables (Uyanık, G. K., & Güler, N. (2013)). We used multiple linear regression to find the influence of macroeconomic, health, race, and education factors on the unemployment rate. The model uses ordinary least squares (OLS) to find the best fit to the data., however, as the number of independent variables increases, the model attempts to account for the variation and the utility of the model might be lost. Second, the t-tests are usually performed when the dataset is small and are primarily used to test the hypothesis. If the p-value resulted from the t-test is less than 0.05 is considered to be significant. It implies that the difference between the two means of variables is significant enough to reject the null hypothesis (Rowntree, D. (1981)). Third, we also used Durbin-Watson measurement to determine the autocorrelation among the consecutive values of the independent variable (Rowntree, D. (1981)). We merged the datasets, checked for missing values, and normalized the values to prepare for the modeling. Once the dataset was pre-processed, it was split into training and testing datasets to avoid overfitting and allow for validation.

## Analysis and Results

### Macroeconomic Factors

We built the two models using the training dataset and validated them using the testing dataset. We found that the accuracy for Linear Regression to be about 70%, while that for Random Forest was 72% . Using the Random Forest model, we calculated relative importance of the features where unemployment insurance came out to be the most important factor.

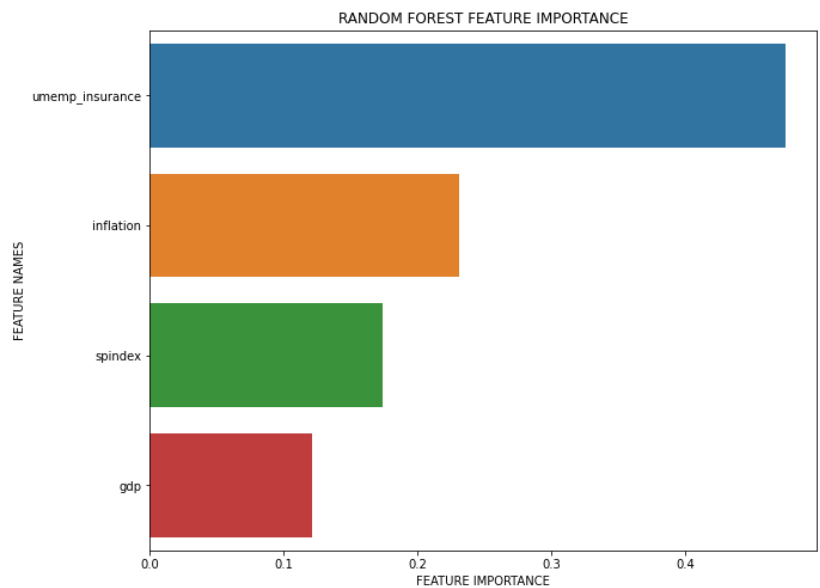

Figure 2 Features Importance Generated by Random Forest Regression Model Trained on Macroeconomic Factors Data Set

From the regression line in Joint Plot and the results from Random Forest Regressor, we can see that unemployment insurance has a positive and strong relationship with unemployment rate as compared to others.

### Microeconomic Factors
### Health

For our study we selected three factors namely Cardiovascular Disease (CVD), Neurological Disorders (ND), and Interpersonal Violence (IV), to investigate their effect on labor market. We applied Random Forest Regression, where the model developed had accuracy score of 91.8%, Mean Square Error (MSE) value of 0.05, and Mean Absolute Error (MAE) value of 0.18. Utilizing the features importance property of model developed using Random Forecast Algorithm, we found out that Interpersonal Violence has a value of 0.46, followed by Neurological Disordered having 0.29 value, with the least contribution by Cardiovascular Diseases that contributed to 0.27 score in developing the model. Figure 3 shows the trends of Cardiovascular Disease (CVD), Neurological Disease, and Interpersonal Violence occurrences per 100,000 persons and unemployment rate. Figure 3 shows the negative relationship between the Cardiovascular Disease (CVD) and Interpersonal Violence with unemployment rate, while it shows a positive relationship with Neurological Disease. Figure 4 on the other hand shows the Features Importance Bar Chart values of which are determined by the trained Random Forest Regression Model.

Contrary to the literature, the negative relationship between CVD and Interpersonal Violence with unemployment rate is noteworthy and confirms other literature, which argued that the relationship between these diseases and unemployment is questionable and "cannot be answered conclusively" (Weber, A., & Lehnert, G. (1997)).

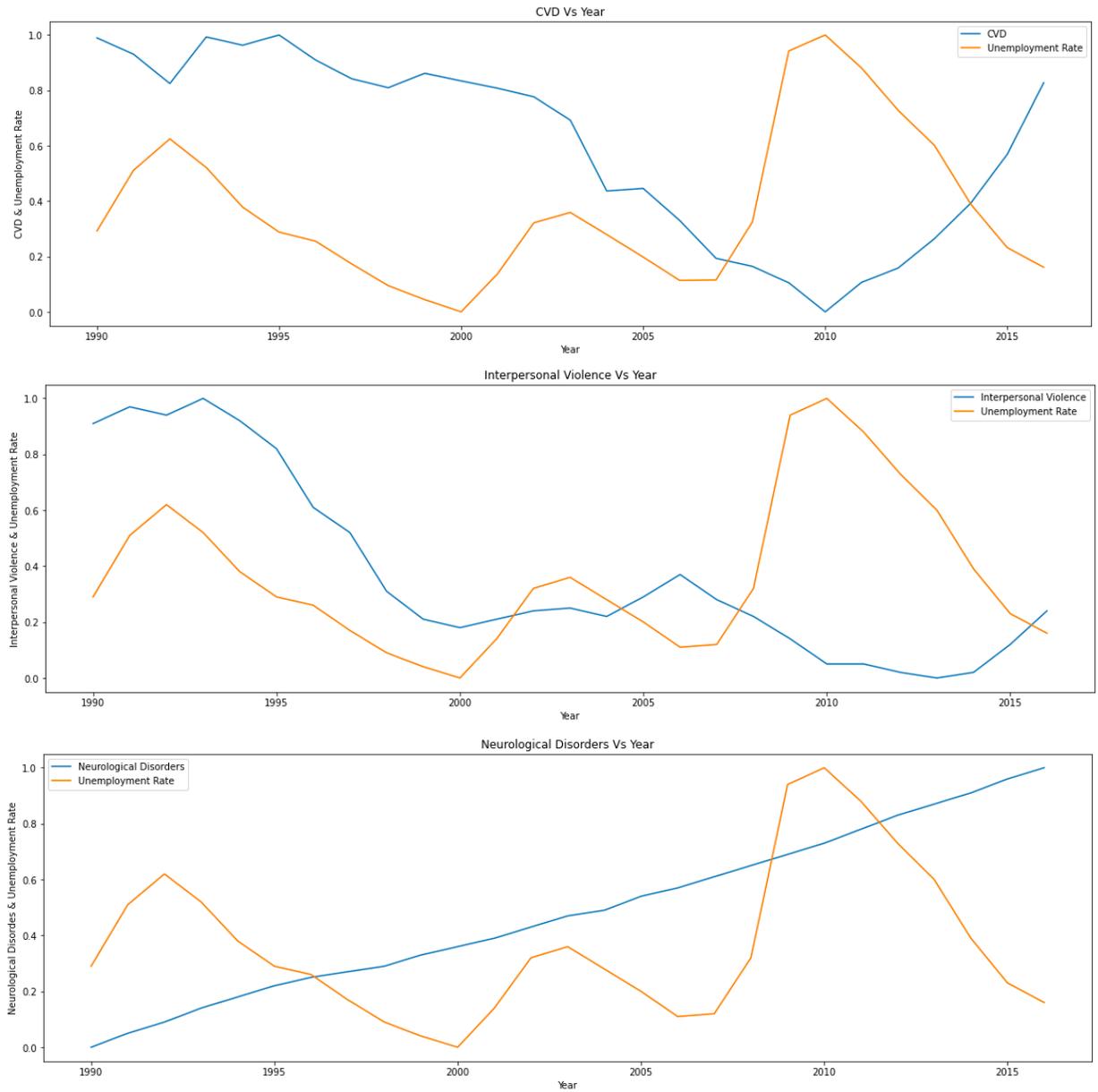

**Figure 3 Trends of Cardiovascular Disease (CVD), Neurological Disease, and Interpersonal Violence occurrences per 100,000 persons and Unemployment Rate**

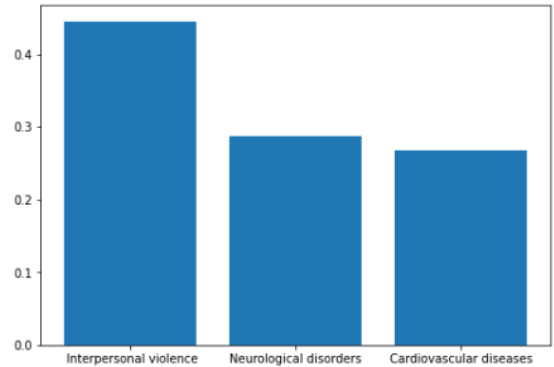

**Figure 4 Features Importance Generated By Random Forest Regression Model Training on Burden of Diseases Data Set**

**Race**

Races which are taken for this study are Asian, Hispanic, Black, and White as they are most populated races in the United States (see Figure 5). We run a random forest regression model to investigate the effect of different races on unemployment rate. Figure 6 shows contribution ratio of the aforementioned ethnicities.

Results show that White contribute highest whereas Asian contribute the lowest in the unemployment rate. The model trained has an R-squared score of 0.90, which means that the model well grasped the variance within the dataset. The individual features importance for White, Black, Hispanic, and Asian are 0.32, 0.28, 0.26, and 0.15 respectively. It can be seen that Asians have higher education rates and better skills hence they are least of the unemployed race.

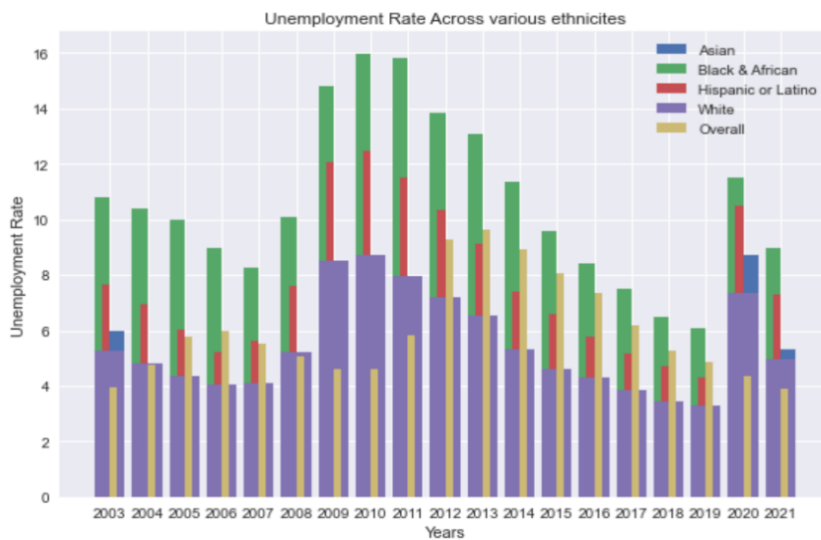

**Figure 5: Unemployment rate across various races.**

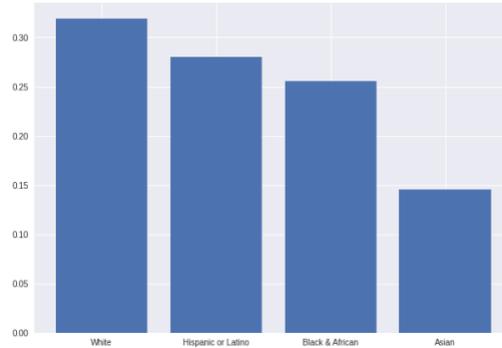

**Figure 6: Features Importance Generated By Random Forest Regression Model Training on Race Data Set**

**Education**

Based on the trend analysis of education level compared to unemployment rates (see Figure 4), we found that people that obtained a higher degree have the lowest unemployment overall, while people with less than a high school diploma have the higher unemployment rates. During the pandemic spike, we found that unemployment rose from roughly 5% to 21% for the less than high school diploma group, which is about 400% increase. As for the group with a bachelor's degree or higher, unemployment rose from roughly 4% to 8.4%, which is about 210% increase. Overall, we can see a clear trend that the higher the education attainment, the lower the employment rate. This agrees with various previous studies such as (Card, 2001; Grossman, 2005; Oreopoulos and Salvanes, 2009; Riddell and song 2011). We used linear regression to understand the effect of the different education levels on the unemployment. The following equation shows the results of the regression:

$$Unempoyment\ Rate = 5.7 + 3.38 * less\ than\ high\ school + 1.84 * high\ school - 0.36 * associate\ degree\ or\ some\ college - 2.26 * Bachelor\ or\ higher$$

The p-value for the independent variables were less than 0.05, which means they can be considered significant in understanding the unemployment rate. Results also showed that R-squared is about 0.73 and the adjusted R-squared is 0.71. We found a positive relationship between high school graduates and who have less than high school, while a negative relationship can be shown for people with college, associate, bachelor, or higher degrees.

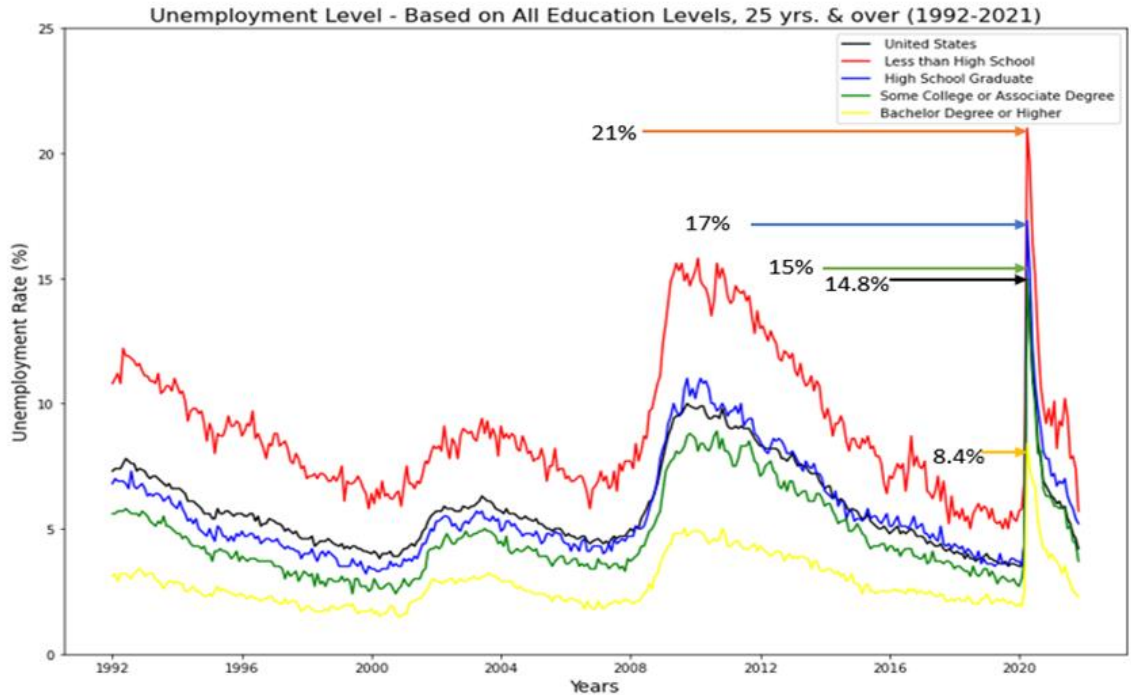

**Figure 7: Education attainment level compared to unemployment**

## Conclusion

While there are a large number of macroeconomic and microeconomic factors affecting the unemployment, we analyzed a few in this study, of which there is significant consensus among the literature. By and large, the fluctuation in economic activity is a determinant factor in unemployment rate. On the macroeconomic level, we found that GDP (negative), inflation (positive), UI (negative), and S&P 500 index (negative) were all significant factors, with inflation being the most important factor. On the microeconomic level, we studied the effect of various health, race, and education factors on unemployment. We found that CVD from the health factor stood out from other two issues (i.e., Interpersonal Violence and Neurological Diseases), to contribute the most in unemployment. This means that with less people contributing in the labor force resulted in less people getting CVD. Similarly, there was a slightly lesser chance of interpersonal violence for people who are unemployed. However, although weak correlation, people having Neurological Disorder are less likely to be employed. Our results vary from the literature we reviewed, where with rise in diseases e.g., CVD and Neurological disorder increases the unemployment, and the impact is stronger for Neurological disorder, similarly, for interpersonal violence, due to increased interpersonal violence, the employment rate lifts up, while we were expecting the same for United States like in Australia, similarly the interpersonal violence related results for United States, however, it was not the case. This confirms that the relationship between health factors and unemployment is still questionable and geographically dependent.

We also found that the racial gaps in unemployment results remain during the different periods with African Americans race being the more unemployed than the rest of the races. Finally, we found that people that obtained a higher degree have the lowest unemployment overall, while people with less than a high school diploma have the higher unemployment rates.

## Funding

No funding was used for this study.

## Ethical Approval

The study was granted exemption from requiring ethics approval as all the data used in this study is publicly available.

## Informed consent

This article does not contain any studies with human participants performed by any of the authors.

## Author's contribution

All authors contributed to the study conception and design. Material preparation, data collection and analysis, as well as the first draft preperation were performed by Alrick Green, Ayesha Nasim, Jaydeep Radadia, Devi Manaswi Kallam, Viswas Kalyanam, and Samfred Owenga. The review of the manuscript was done by Huthaifa I. Ashqar and all authors commented on previous versions of the manuscript. All authors read and approved the final manuscript.

## Conflict of interest

The authors have no competing interests to declare that are relevant to the content of this article.

## Data availability statement

The data was collected from United Sates federal website: https://www.bls.gov/.

## Acknowledgement

We acknowledge the U.S. Bureau Of Labor Statistics for providing the dataset.